\documentclass[
    trackchanges, 
    astrosymb, 
    tighten, 
    twocolumn, 
]{aastex631}
\usepackage{microtype}
\usepackage{amsmath}
\usepackage{institutions}

\def\homogeneous{\texttt{homogeneous}}
\def\xione{\texttt{xi-1}}
\def\xiten{\texttt{xi-10}}
\def\xihundred{\texttt{xi-100}}
\def\zetahalf{\texttt{zeta-0.5}}
\def\zetatwo{\texttt{zeta-2}}

\begin{document}

\title{Modeling the Far-Infrared Polarization Spectrum of a High-Mass Star Forming Cloud}

\author[0000-0002-3455-1826]{Dennis Lee}
    \affiliation{\CIERA}
    \affiliation{\NU}
\author[0000-0002-9209-7916]{Che-Yu Chen}
    \affiliation{\LLNL}
\author[0000-0003-1288-2656]{Giles Novak}
    \affiliation{\CIERA}
    \affiliation{\NU}
\author[0000-0003-0016-0533]{David T. Chuss}
    \affiliation{\Villanova}
\author[0000-0002-5216-8062]{Erin G. Cox}
    \affiliation{\CIERA}
\author[0009-0006-4830-163X]{Kaitlyn Karpovich}
    \affiliation{\Villanova}
\author{Peter Ashton}
    \affiliation{\SOFIAUSRA}
\author{Marc Berthoud}
    \affiliation{\Uchicagoengineering}
    \affiliation{\CIERA}
\author{Zhi-Yun Li}
    \affiliation{\UVA}
\author[0000-0003-3503-3446]{Joseph M. Michail}
    \affiliation{\CIERA}
    \affiliation{\NU}

\begin{abstract}
The polarization spectrum, or wavelength dependence of the polarization fraction, of interstellar dust emission provides important insights into the grain alignment mechanism of interstellar dust grains. 
We investigate the far-infrared polarization spectrum of a realistic simulated high-mass star forming cloud under various models of grain alignment and emission. 
We find that neither a homogeneous grain alignment model nor a grain alignment model that includes collisional dealignment is able to produce the falling spectrum seen in observations. 
On the other hand, we find that a grain alignment model with grain alignment efficiency dependent on local temperature is capable of producing a falling spectrum that is in qualitative agreement with observations of OMC-1.
For the model most in agreement with OMC-1, we find no correlation between temperature and the slope of the polarization spectrum.
However, we do find a positive correlation between column density and the slope of the polarization spectrum. 
We suggest this latter correlation to be the result of wavelength-dependent polarization by absorption.
\end{abstract}


\section{Introduction} \label{sec:intro}

While observational studies of magnetic fields in molecular clouds are difficult, they are crucial for addressing the many open questions on their role in star formation \citep{2012ARA&A..50...29C,2014prpl.conf..101L, 2019FrASS...6...15P}. 
The most accessible way to carry these out is to observe the polarized emission from molecular clouds at far-infrared to millimeter wavelengths \citep{2000PASP..112.1215H}. 
These measurements trace the thermal emission from magnetically aligned dust grains in the molecular clouds. 
Radiative alignment torques (RATs) is the leading theory for magnetic grain alignment \citep{1976Ap&SS..43..291D, 1997ApJ...480..633D, 2007MNRAS.378..910L, 2015ARA&A..53..501A}. In RATs theory, anisotropic optical/near-infrared radiation fields align irregularly shaped dust grains via the transfer of angular momentum from the radiation field to the dust grains.
The end result of this process is that the grains spin about their axis of greatest moment of inertia, which are preferentially aligned with the external magnetic field.
The resulting polarized emission is oriented perpendicular to the plane-of-sky projection of the magnetic field.

Using this polarized emission, numerous statistical methods have been developed to infer physical properties from observations of the magnetic field in these star-forming regions. 
Techniques to estimate field strength and other properties, for example, include 
{the Davis-Chandrasekhar-Fermi (DCF) method (\citealt{Davis_1951,CF_1953}; also see e.g.,~\citealt{OSG2001, Hildebrand_09, PSLi_2021, Chen_DCF_2022}), the DCF-related polarization dispersion analyses \citep[PDA; e.g.,][]{2009ApJ...706.1504H, 2016ApJ...820...38H, 2019ApJ...872..187C}, and histograms of relative orientations \citep[HRO; e.g.,][]{Soler_HRO_2013,2016A&A...586A.138P, 2016ApJ...829...84C, 2019A&A...629A..96S, 2021ApJ...918...39L}.}
As a result, understanding the physics of the grain alignment mechanism is essential to ensuring proper interpretation of polarization measurement derived analyses \citep{2015ARA&A..53..501A}.

One avenue for investigating grain alignment is to study the wavelength dependence of the polarization fraction in a molecular cloud \citep{1999ApJ...516..834H}. This \textit{polarization spectrum} is created by observing the polarization fraction for a sight line as a function of wavelength ($p$ vs $\lambda$). 
The extent to which the properties of the polarization spectrum (e.g., its slope) depend on environmental conditions can potentially place constraints on the nature of the grain alignment physics \citep{2018ApJ...857...10A, 2021ApJ...907...46M}.


In molecular clouds, the far-infrared ($\sim$50 to $\sim$300~\micron) polarization spectrum has been observed to be falling \citep[lower polarization fraction at longer wavelengths;][]{1999ApJ...516..834H,2008ApJ...679L..25V,2013ApJ...773...29Z, 2021ApJ...907...46M}. \citet{1999ApJ...516..834H} attributed this behavior to a specific effect arising in heterogeneous clouds. \citet{2021ApJ...907...46M} refers to this as the heterogeneous cloud effect (HCE). 
The HCE describes the scenario where regions with warmer grains and regions with colder grains are both present along a sight line. Grains in the warmer regions have a relatively higher grain alignment efficiency and thus emit with a higher polarization fraction. Meanwhile, grains in cooler regions have a lower grain alignment efficiency and emit with a correspondingly lower polarization fraction. As the warmer grains contribute a larger fraction of intensity at shorter wavelengths and the colder, poorly-aligned grains contribute more radiation at longer wavelengths, the polarization fraction is observed to fall with wavelength. This results in the negatively sloped $p$ vs $\lambda$ behavior that has been observed \citep{1999ApJ...516..834H,2008ApJ...679L..25V,2013ApJ...773...29Z, 2021ApJ...907...46M}. 

The temperature-dependent grain alignment efficiency can be naturally explained by RATs.
Grains in the warmer regions are exposed to a more intense anisotropic radiation field and hence are also expected to be more well aligned. On the other hand, the grains of colder, denser regions are shielded from the radiation and are, as a result, less efficiently aligned.

Nonetheless, the falling polarization spectrum need not be explained by RATs or indeed HCE at all. In principle, it is possible that variations in the magnetic field direction within beam volume alone (so-called ``field tangling'') can reduce the observed polarization fraction and hence be the source of the falling spectrum. Even assuming a homogeneous grain alignment model, one expects a falling spectrum if the colder regions along a sight line suffer from more field tangling and cancellation than the warmer regions. 
The gravitational collapse induced by nearby star formation is one possible scenario for increased field tangling and cancellation in denser and colder regions.

Even if the HCE does operate in molecular clouds, it is possible that the alignment mechanism is actually regulated by volume density rather than directly by temperature \citep[e.g.,][]{2013A&A...559A.133Y}. 
More collisions in denser environments, for example, can ``unalign'' grains, resulting in lower observed polarization fractions \citep{2015ARA&A..53..501A}. 

OMC-1, located at a distance of 390 pc as part of the Orion Nebula complex, is one of the nearest sites of massive star formation \citep{2017ApJ...834..142K}. 
Using the HAWC\texttt{+} instrument on the Stratospheric Observatory for Infrared Astronomy \citep[SOFIA;][]{2018JAI.....740008H}, \citet{2021ApJ...907...46M} studied the far-infrared polarization spectrum of OMC-1. Finding an overall falling polarization spectrum, \citet{2021ApJ...907...46M} attributed this to the HCE. The authors argued that this result, taken together with correlations seen between the slope of the polarization spectrum and local environmental conditions, provide evidence for RATs operating in OMC-1. 

In this work, we aim to test these conclusions via synthetic observations of a realistic, heterogeneous cloud simulation using various grain alignment models. We compute the polarization spectra resulting from these various grain alignment models and compare to the OMC-1 observations.
The paper is organized as follows: Section~\ref{sec:methods} reviews the polarization spectrum observations of OMC-1 with SOFIA/HAWC\texttt{+} and introduces our simulation and synthetic observations. In particular, we discuss the grain alignment models we consider in Section~\ref{sec:methods:grain_alignment_prescriptions}. Section~\ref{sec:analysisresults} presents the polarization spectrum analysis of our synthetic simulations and compares the results with the analogous analysis of the OMC-1 observations. Section~\ref{sec:discussion} discusses the implications of the analysis. We present concluding thoughts in Section~\ref{sec:conclusison}.

\section{Methods}\label{sec:methods}
\subsection{Polarimetric Observations of OMC-1}\label{subsec:observations}
The OMC-1 region was observed using SOFIA/HAWC\texttt{+} in four bands centered at 53, 89, 154, and 214~\micron~with resolutions of 5\arcsec, 8\arcsec, 14\arcsec, and 19\arcsec~respectively. The polarimetric observations were conducted between December~2016 and December~2018. Detailed descriptions on the observations and data reduction are presented in \citet{2019ApJ...872..187C}. 

\citet{2021ApJ...907...46M} studied the polarization properties of OMC-1 across all four bands using these observations, but adopted slightly lower resolutions (20.5\arcsec at 214~\micron) to account for additional smoothing during data reduction. In order to compare across all four bands, \citet{2021ApJ...907...46M} smoothed all data to 20.5\arcsec~to match the lowest resolution data. In addition to typical cuts based on signal-to-noise thresholds \citep[described in detail in][]{2019ApJ...872..187C}, \citet{2021ApJ...907...46M} also excludes regions where the variation in polarization angles between wavelengths is greater than $15^\circ$. We discuss the impact of this in Section~\ref{sec:methods:synthetic_observations}.

\subsection{Numerical Simulation}\label{sec:methods:simulations} 
We considered a radiation-magnetohydrodynamic (RMHD) simulation that follows the formation of massive stars in a magnetized, turbulent cloud using the \textsc{Orion2} code \citep{ORION2}. 
The simulation was initialized as a dense cloud core of mass $M_{\rm core} = 10^3$\,M$_\odot$ with a power-law density profile $\rho(r) \propto r^{-3/2}$ out to the radius of the core $R_{\rm core} = 0.41$\,pc. 
The mean column density is therefore $\Sigma \approx 0.4$\,g$\,$cm$^{-2}$ ($N_{\text{H}_2}\approx10^{23}$$\,$cm$^{-2}$), consistent with the typical range of $\Sigma$ values ($0.1-1$\,g$\,$cm$^{-2}$) in infrared dark clouds (IRDCs) where high-mass star formation occurs \citep[e.g., see the review by][]{Motte_HMSF_2018}.
A turbulent velocity field was added to the core with Mach number ${\cal M} = 2.43$, which corresponds to a virial parameter $\alpha_{\rm vir} = 3.73$. The initially uniform magnetic field strength is chosen so that the normalized mass-to-flux ratio is $3.0$ inside the core ($r < R_{\rm core}$).
Three levels of adaptive mesh refinement (AMR) were applied so that the highest resolution is $\Delta x \approx 164$\,AU.
{For details on the treatment and application of ideal MHD with AMR in \textsc{Orion2}, see \cite{PSLi_MHDcode_2012, PSLi_MHD_2015}.}

The initial gas temperature inside the core is $T_{\rm gas} = 35$\,K.  This value is also set to be the temperature floor of the simulation in order to avoid tiny and/or negative temperatures from numerical rarefaction. 
During the formation and evolution of the protostar, radiation transport is handled by a frequency-integrated flux limited diffusion (FLD) approximation with dust opacity models from \cite{Semenov_opacity_2003}.
More details about the protostellar model and optical properties adopted in \textsc{Orion2} can be found in \cite{Offner_2009} and \cite{AJC_2011}.
{Note that} our simulation setup is {almost identical} to that described in \cite{AJC_2011}, but with the addition of the magnetic field {to include more complete physics}. 

\begin{figure*}
    \centering
    \includegraphics[width=0.98\textwidth]{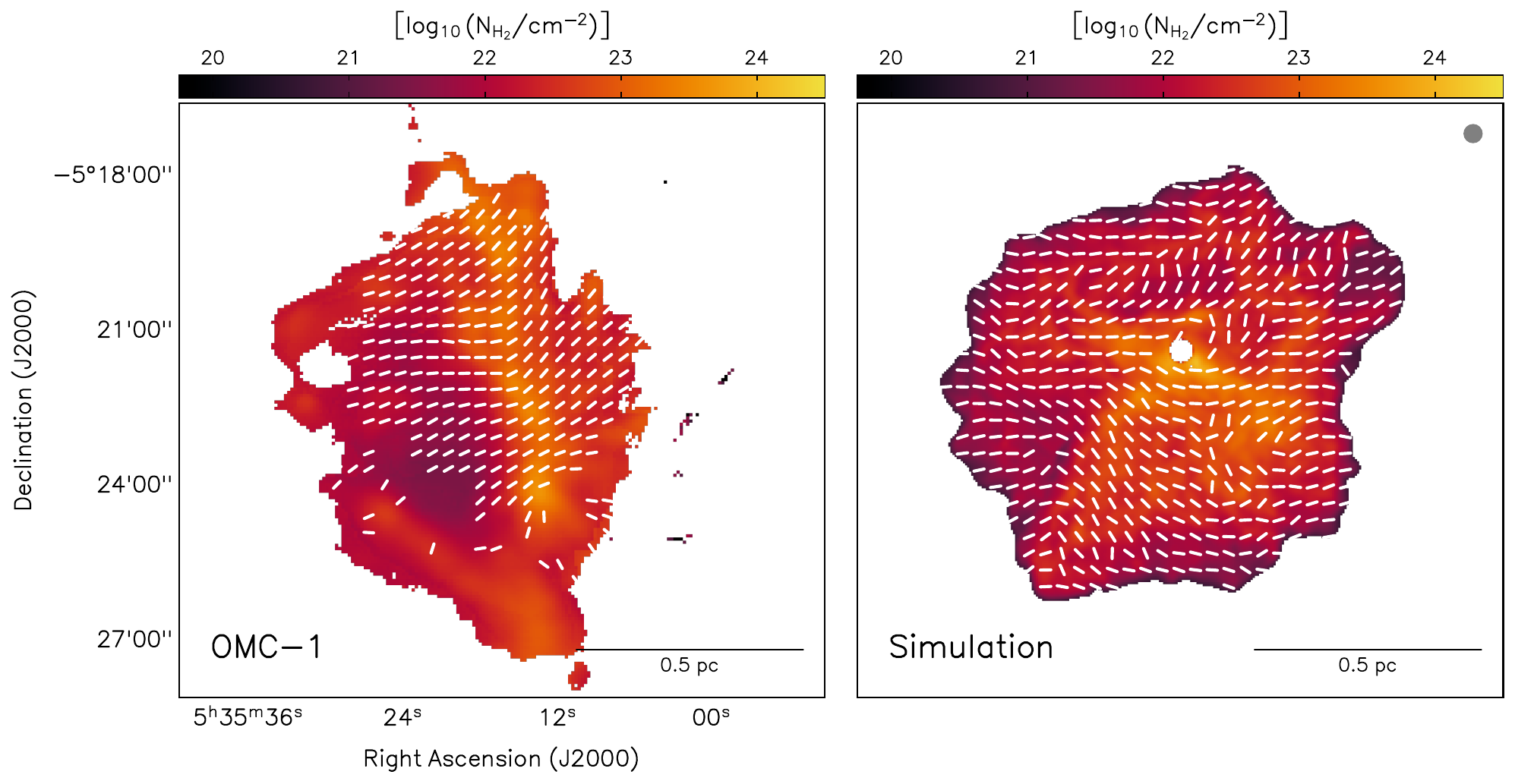}
    \caption{
    Maps of column density in molecular hydrogen ($N_{\text{H}_2}$) for OMC-1 (left) and our fiducial simulation, \xiten~(right). 
    Column density values for OMC-1 are as computed by \citet{2019ApJ...872..187C} and used in \citet{2021ApJ...907...46M}. 
    Column density values for our simulations were obtained via the spectral energy distribution (SED) fitting detailed in Section~\ref{sec:methods:synthetic_observations}. 
    Line segments shown in white represent the magnetic field direction inferred from the 214~\micron~polarized emission.
    All simulation data is convolved to 20.5\arcsec~in order to match the resolution of the OMC-1 observations (indicated by the beam in the upper right).  
    Line segments are plotted every 20.5\arcsec. 
    Our simulated cloud is similar in physical extent to OMC-1, as indicated by the scale bar at the lower right of each panel. 
    Only the \xiten~model is shown here, but all the SED-derived column density and temperature maps from the various models are similar. \label{fig:sim_vs_coldense_spatial}}
\end{figure*}

For our synthetic observation, we consider a temporal snapshot when roughly $3.5\%$ of the cloud core mass is inside the forming protostars (two massive protostars with $\sim 21$ and $\sim $14\,M$_\odot$). At this evolutionary stage the gas material is not overly centered around the protostars, providing a physical environment comparable to OMC-1 (see Section~\ref{sec:methods:synthetic_observations}). 
{A sample synthetic polarization map of our simulation is shown in Figure~\ref{fig:sim_vs_coldense_spatial} (right panel).}

\subsection{Grain Alignment and Polarization Prescriptions}
\label{sec:methods:grain_alignment_prescriptions}

We use the approach described in \cite{Lam_2021} and \cite{Yang2021} to numerically calculate the polarized emission from simulations.
This method---in contrast to fully solving the complicated radiative transfer process (e.g., \textsc{Polaris}; \citealt{2016A&A...593A..87R})---allows us to test various parameters one at a time to investigate the critical factors in the process of polarized emission.
While the full details of this approach can be found in \cite{Lam_2021} and \cite{Yang2021}, we briefly describe it here.

The fundamental criterion for magnetic alignment of spinning dust grains is that the grains have to gyrate quickly around the magnetic field before the dust-gas collisions disrupt this process.
Consider the ratio between the Larmor precession timescale $t_L$ and the gas damping timescale $t_d$ and assume that the dust grains and gas particles are in thermal equilibrium ($T_{\rm dust} = T_{\rm gas} = T$). 
In this case, the magnetic alignment criterion is simply

\begin{align}
    \frac{t_L}{t_d} = 0.77 \times \left(\frac{a_{\rm mm}}{\hat\chi}\right) &\left( \frac{B}{100~\mu\mathrm{G}} \right)^{-1} \notag \\
    & \left(\frac{n_{\rm g}}{ 10^5~\mathrm{cm^{-3}}}\right)  \left(\frac{T}{15~\mathrm{K}}\right)^{3/2} < \frac{1}{\eta},
    \label{eq:ratio}
\end{align}
or
\begin{align}
    \label{eq:aligncrit}
    \xi & \equiv \frac{\hat\chi/\eta}{a_\mathrm{mm}} \notag\\
    &> 0.77 \times \left( \frac{B}{100\,\mu\mathrm{G}} \right)^{-1} \left( \frac{n_\mathrm{g}} {10^5\,\mathrm{cm^{-3}}} \right) \left( \frac{T}{15\,\mathrm{K}} \right)^{3/2},
\end{align}
where $a_{\rm mm}$ is the radius of individual grains in units of millimeter, $\hat\chi$ a dimensionless magnetic susceptibility determined by the composition of dust grains {(see \citealt{Yang2021} for more details)}, $B$ the magnetic field strength, $n_{\rm g}$ the number density of the gas, and $\eta\ (>1)$ the number of gyrations per gas damping time needed for grain alignment.
The value of $\xi$ thus serves as an alignment parameter that can be used to switch between various grain alignment prescriptions\footnote{We note that there is degeneracy in the value of $\xi$ between $\hat\chi$ and $a_{\rm mm}$. While the exact value of $\eta$ is uncertain, \cite{Yang2021} suggested a fiducial value of $\eta \sim 10$.} \citep[see e.g.,][]{Lam_2021}.

The polarized emission can then be derived following the method described in \cite{Lam_2021} by solving the vector radiation transfer equation for the Stokes vector $\mathbf{S} = (I, Q, U)$: 
\begin{equation}
    \frac{1}{\rho}\frac{d}{ds} \mathbf{S} = -\mathcal{K}\,\mathbf{S} + B_\nu(T)\,\mathbf{a},
    \label{eq:vrt}
\end{equation}
where $\rho$ is the gas density, $s$ the distance along the sight line, and $B_\nu (T)$ the Planck function.
The extinction matrix $\mathcal{K}$ and emission vector $\mathbf{a}$ are related to the extinction and absorption opacities, which are dependent on the magnetic field structure, the alignment criterion (Equation~\ref{eq:aligncrit}), the dust opacity $\kappa_\nu$, and the polarizability parameter $\alpha_p$ (related to the maximum degree of polarization $p_0$ as $p_0 \approx \alpha_p / (1 - \alpha_p/6)$). {See Appendix B of \cite{Lam_2021} for full discussions.}

The formal solution of Eq.\,(\ref{eq:vrt}) is 
\begin{equation}
    \mathbf{S} = \int \mathcal{T}(s)\,\mathbf{a} \,B_\nu(T)\,\rho\,ds,
    \label{eq:FormalSolution}
\end{equation}
where the matrix $\mathcal{T}(s)$ is obtained from  the integral
 \begin{equation}
     \mathcal{T}(s) = \Pi_s^\infty e^{-\rho(s')\mathcal{K}(s')ds'}.
     \label{eq:Tau}
 \end{equation}
Note that $\Pi$ denotes the order-preserved geometric integration, because the extinction matrix may not be commutative. 
We refer the reader to Appendix~B of \cite{Lam_2021} for the complete calculation of $\mathbf{S}$, we would like to point out that in the simplest case of optically thin dust with homogeneous ($\xi \rightarrow\infty$, constant $\alpha_p$) grain alignment, Equation~\ref{eq:FormalSolution} reduces to the commonly-adopted equations fir the Stokes parameters \citep{Fiege_Pudritz_2000}:
\begin{subequations}
\begin{align}
    I &= \int \rho \left( 1 - \alpha_p \left( \frac{\cos^2 \gamma}{2} - \frac{1}{3} \right) \right) ds,\\
    Q &= \alpha_p \int \rho\,\cos 2\psi\,\cos^2 \gamma\,ds,\\
    U &= \alpha_p \int \rho\,\sin 2\psi\,\cos^2 \gamma\,ds,
\end{align}
\end{subequations}
where $\psi$ is the angle between the magnetic field and the direction of positive $Q$ in the sky plane and $\gamma$ is the inclination angle of the magnetic field relative to the plane of the sky.

We consider three classes of models that differ by the grain alignment properties as constrained by the grain alignment parameters, $\xi$ and $\alpha_p$:

\begin{enumerate}

    \item \textbf{Homogeneous}: As our simplest model, we consider the scenario of perfect grain alignment. With $\xi = \infty$ and  constant $\alpha_p = 0.1$ \citep[a typical polarization level at cloud scales; e.g., see][]{2019FrASS...6...15P, 2023ASPC..534..193P}, grains throughout the simulated cloud, regardless of local conditions, are considered magnetically aligned and emit with $10\%$ polarization fraction. The derived polarization thus depends only on the magnetic field structure as well as the optical depth effects.
        
    \item \textbf{Collisional Depolarization}: These models are based on \cite{Lam_2021} for selected values of $\xi$ {(see Equation~\ref{eq:aligncrit} for the definition)}. 
    {By varying $\xi$, thus adjusting the criterion  for alignment, these models simulate the depolarization effect due to collisions between dust grains and gas particles.
    A lower value of $\xi$ corresponds to a stricter grain alignment criterion.
    For each voxel in the simulation, grains are considered magnetically aligned if the criterion in Equation~\ref{eq:aligncrit} is satisfied (i.e., Equation 10 of \citealt{Lam_2021}).
   In addition, effects from the homogeneous model, which are related to magnetic field geometry and optical depth, remain present.}
    We consider three values for $\xi$: 1, 10, and 100 (\xione, \xiten, \xihundred).
    We adopt $\xi=10$ (\xiten) as our fiducial model.
    

    \item \textbf{Temperature-Dependent~Polarizability}: We extend the collisional depolarization model by adopting a power-law temperature dependence in the polarizability parameter, $\alpha_p \equiv 0.1\,(T/35\,{\rm K})^\zeta$ (with a maximum value set such that $\alpha_p \le 0.9$)\footnote{In principle, this threshold can result in a nonphysical polarization fraction of $p>1$ when $\alpha_p = 0.9$. However, this requires $T>320\,$K which occurs in an insignificant number of pixels  ($0.009\%$).}. This setting allows warmer regions to have a higher degree of magnetic alignment. 
    This is intended to model to the effect of RATs whereby warmer, and hence faster-spinning grains are better aligned with the magnetic field.
    We consider two different values for the power-law index $\zeta$: 0.5 and 2.0.
    \zetahalf\,($\alpha_p \propto T^{0.5}$) models a weaker dependence on temperature while \zetatwo~($\alpha_p \propto T^2$) models a stronger dependence on temperature. 
\end{enumerate}


\subsection{Synthetic Observations}
\label{sec:methods:synthetic_observations}

For each grain alignment prescription, we generate synthetic observations of the simulated cloud at each HAWC\texttt{+} band: 53, 89, 154, and 214~\micron. 
Each set of synthetic observations is smoothed to the common resolution of 20.5\arcsec to match the observations ($\approx 0.04$\,pc at $d=400$\,pc). 
{We select our fiducial viewing angle as $45\,^\circ$ from both $x$- and $z$-axes such that the background magnetic field (initially along $z$-axis) is oriented neither along the line of sight nor the plane of sky (see e.g.,~\citealt{Chen_Bangle_2019} for more discussions on viewing angle and polarization observations).}

We compute the polarization fraction ($p$) and polarization angle ($\varphi$) from the Stokes parameters as:
\begin{equation}
     p  = \frac{\sqrt{Q^2+U^2}}{I} \label{eq:polfrac}
\end{equation}
\begin{equation}
    \varphi = \frac{1}{2} \arctan \left( \frac{U}{Q} \right) \label{eq:polangle}
\end{equation}
We mask regions of extremely low polarization that are not robustly detectable by observations ($p < 0.002$). This accounts for a small fraction of our simulation data ($<0.1\%$). 


The synthetic observations ($I_\nu$) for each wavelength ($\lambda$) are then fit to a modified blackbody function for each sight line to obtain the column density ($N_{\text{H}_2}$) and temperature ($T$):
\begin{equation}
I_\nu \left(\lambda\right) =\left(1 - e^{- \tau_\lambda}\right) B_\nu\left(T\right),
\end{equation}
where the optical depth $\tau_\lambda$ is:
\begin{equation}
    \tau_\lambda = N_{\text{H}_2} \kappa_\lambda \mu_{\text{H}_2} m_{\text{H}} 
\end{equation}

We use $\mu_{\text{H}_2}=2.8$ and $m_{\text{H}}=1.6737 \times 10^{-24}$ grams. For $\kappa_{\lambda}$, we calculate interpolated values of the wavelength-dependent $K_{\text{abs}}$ from \citet{2001ApJ...548..296W} {and \citet{2001ApJ...554..778L}}---the same model used to generate the synthetic observations---which are then modified by a gas-to-dust ratio of 100 (i.e, $\kappa_{\lambda} = K_{\text{abs}} \times 0.01~\text{cm}^2/\text{g}$).

The resulting column density and temperature values are then used for the rest of this analysis. We mask diffuse regions ($N_{\text{H}_2} < 5 \times 10^{20}$ cm$^{-2}$) that are typically on the periphery of the cloud. We also mask regions that are not well fit by the modified blackbody function (e.g., sight lines that do not converge). These are largely centered on the emission peak in the center of the map (see Figure~\ref{fig:sim_vs_coldense_spatial}). Together, this accounts for approximately $\sim6\%$ of the data. We also verify that the column density values obtained through the spectral energy distribution (SED) fitting procedure (median $N_{{\rm H}_2}=10^{22.2}$ cm$^{-2}$) are consistent with the values computed from directly integrating the simulations along each sight line (median $N_{{\rm H}_2}=10^{22.3}$ cm$^{-2}$).

Unlike \citet{2021ApJ...907...46M}, we do not remove sight lines where the variation in polarization angles between each band is greater than $15^\circ$. 
This cut was done in \citet{2021ApJ...907...46M} in order to focus the analysis on polarization spectrum behavior due to dust grain properties rather than field geometry.
In this work, we include sight lines with variation in polarization angles so that we can consider the possibility of field geometry as contributing to the behavior of the polarization spectrum.

\begin{figure}
    \centering
    \includegraphics[width=0.45\textwidth]{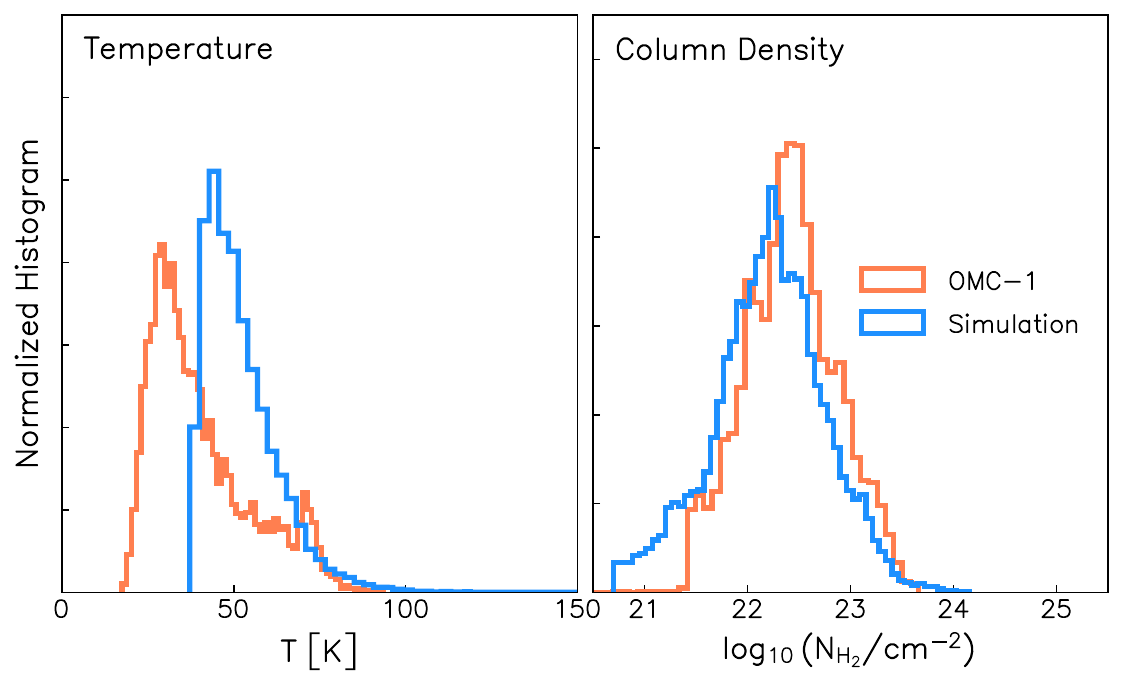}
    \caption{Temperature distribution (left) and column density distribution (right) of OMC-1 (orange) and our synthetic observations (blue) respectively. Temperature and column density values for the synthetic simulations were obtained via the spectral energy distribution (SED) fitting detailed in Section~\ref{sec:methods:synthetic_observations}.
    While our simulated cloud is overall warmer than OMC-1, both our simulated cloud and OMC-1 show similar column density distributions.
    }
    \label{fig:hist_comparison}
\end{figure}

While our simulation setup was originally designed to study high mass star formation processes and thus was not aimed at representing any specific star forming cloud in particular, overall, we find that our simulated cloud is a reasonable analog for OMC-1. As shown in Figure~\ref{fig:sim_vs_coldense_spatial}, the simulated cloud has a similar physical size as OMC-1 (assuming a distance of $\sim390$~pc). The inferred magnetic field direction derived from the 214~\micron~observations of OMC-1 and the simulations are overlaid in Figure~\ref{fig:sim_vs_coldense_spatial}.
The circular standard deviation of the inferred field direction for the observations and the simulations are not too different at $21\,^\circ$ and $35\,^\circ$, respectively.
Figure~\ref{fig:hist_comparison} shows the column density and temperature distributions of OMC-1 compared to our simulation. Both our simulated cloud and OMC-1 show similar column density distributions with median of $\log_{10}\left(N_{{\rm H}_2}/{\rm cm}^{-2}\right)\approx22$ ($22.2$ and $22.4$ for the simulation and OMC-1 respectively). The temperature distribution of our {simulated cloud} is somewhat higher ($\sim 1.4\times$). This {is likely, at least in part, due to the $35$\,K initial gas temperature and temperature floor of our simulation (see Section~\ref{sec:methods:simulations})}. Finally, \citet{2021ApJ...907...46M} report that about 25\% of their sight lines have $\tau_{53} \geq 0.5$. For our synthetic sight lines, we report 28\% of sight lines with $\tau_{53} \geq 0.5$ in OMC-1. In summary, despite some modest differences, our simulation provides a suitable proxy for studying the physics of OMC-1.


\section{Analysis and Results}\label{sec:analysisresults}
\subsection{Global Polarization Spectrum}\label{sec:analysisresults:global_spectrum}
\begin{figure*}
    \centering
    \includegraphics[width=0.96\textwidth]{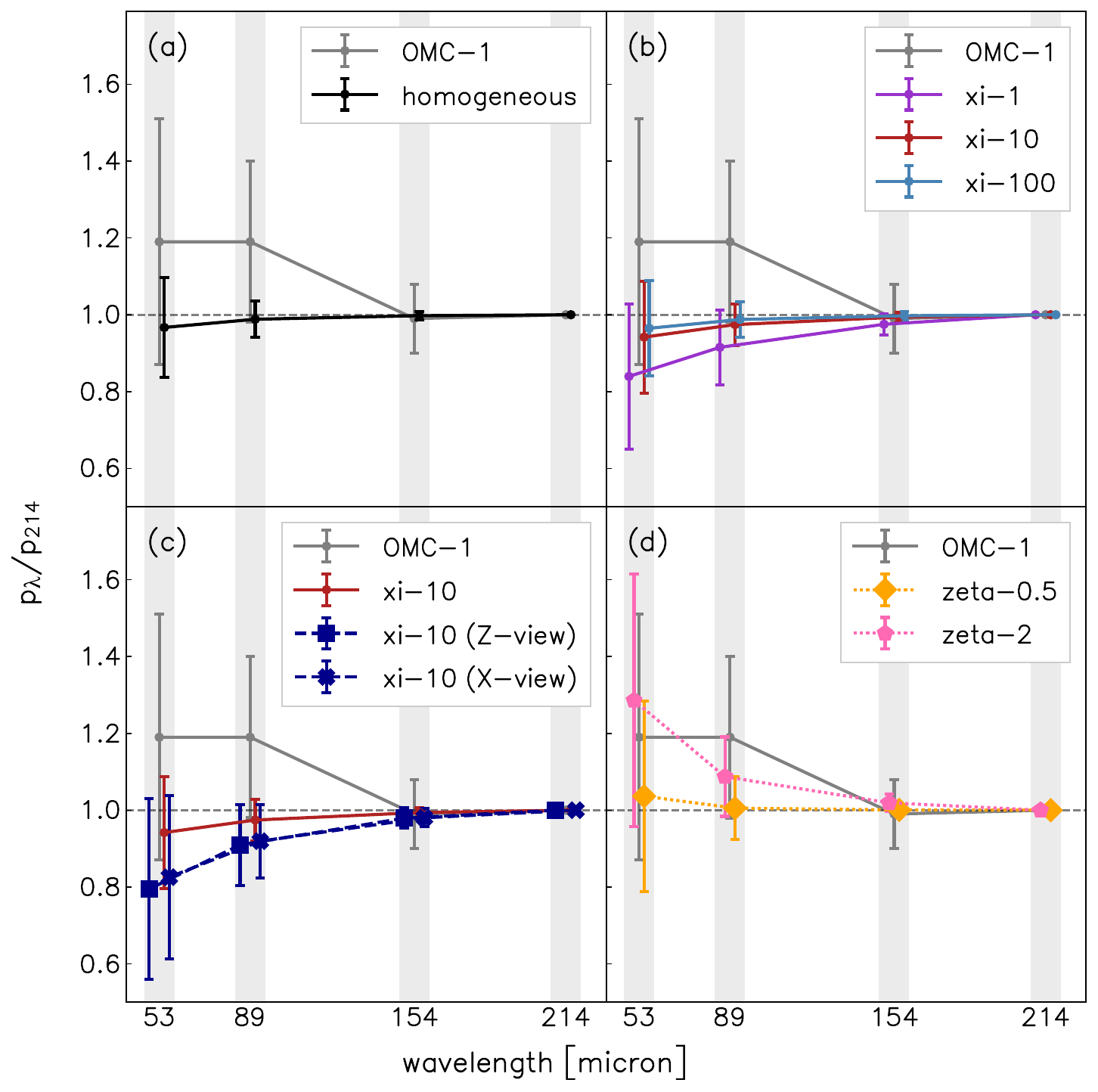}
    \caption{Global polarization spectrum for each set of synthetic observations (see Section~\ref{sec:methods:grain_alignment_prescriptions}). For better visibility, models are plotted with a slight offset at each HAWC\texttt{+} band. The polarization spectrum computed for OMC-1 \citep{2021ApJ...907...46M} is shown for comparison in grey. 
    \textit{(a)}:~Homogeneous grain alignment model (\homogeneous). 
    \textit{(b)}: Collisional Depolarization grain alignment models varying the $\xi$ parameter: 1, 10, and 100 (\xione, \xiten, \xihundred). 
    \textit{(c)}: Our fiducial collisional depolarization grain alignment model (\xiten) for the nominal as well as two other viewing angles. 
    \textit{(d)}: Temperature-Dependent Polarizability models where the polarizability is dependent on the temperature (\zetahalf, \zetatwo). 
    The only model in this work that is able to produce a falling polarization spectrum comparable to OMC-1 is the \zetatwo~temperature-dependent polarizability model shown in panel (d).
    \label{fig:all_polspec}}
\end{figure*}

For each set of synthetic observation{s}, we produce the global polarization spectrum by finding the median value of the polarization fraction ratio $\left(p_\lambda / p_\text{214}\right)$ across the cloud.
The associated median absolute deviation (MAD) values are also calculated. These are shown in Figure~\ref{fig:all_polspec} and Table \ref{tab:global_polspec}, which also include the analogous results obtained for OMC-1 by \citet{2021ApJ...907...46M}. 

The homogeneous grain alignment model (\homogeneous) produces a relatively flat global polarization spectrum (Figure~\ref{fig:all_polspec}a) indicating consistent polarization fractions across the four bands. 

The global polarization spectrum of models using the collisional depolarization grain alignment model (\xione, \xiten, \xihundred) show flat or increasing polarization spectrum (Figure~\ref{fig:all_polspec}b). 
Varying the alignment parameter $\xi$ in our grain alignment model changes the spectrum shape from rising at low values of $\xi$ to a flatter shape at high values of $\xi$ (Figure~\ref{fig:all_polspec}b).
As expected, at large $\xi$ value, the grain alignment model approaches the model of uniform grain alignment ($\xi=\infty$). 
None of these collisional depolarization grain alignment models (Figure~\ref{fig:all_polspec}b)---regardless of $\xi$-----produced a falling polarization spectrum.

Figure~\ref{fig:all_polspec}c shows the fiducial collisional depolarization model (\xiten) in the nominal viewing angle along with two different viewing angles. The \texttt{Z-view} model represents the viewing angle along the initial direction of the magnetic field. The \texttt{X-view} model represents a viewing angle where the magnetic field is orthogonal to the line of sight. 
Despite the different projected magnetic field directions when viewed from different angles, analysis of the synthetic observations along all three viewing angles result in a rising polarization spectrum (i.e, positive slope).
This suggests that the projected magnetic field morphology is not the key parameter contributing to the wavelength-dependent polarization fraction.


Extending our investigations to test the concept of RATs, we considered two models where $\alpha_p$ is dependent on temperature ({temperature-dependent~polarizability} models; \zetahalf, \zetatwo). 
By adding a simple power-law dependence on temperature to the polarization coefficient $\alpha_p (T)\equiv 0.1(T/35\,\text{K})^\zeta$ (see Section~\ref{sec:methods:grain_alignment_prescriptions}), our fiducial model (\xiten) changed from a rising polarization spectrum to a falling spectrum. 
The results are shown in Figure~\ref{fig:all_polspec}d which also indicate that the model with the stronger the dependence on temperature (\zetatwo) has the more steeply falling overall spectrum comparable to the polarization spectrum observed in OMC-1.
We discuss the possible origins of the falling polarization spectrum in detail in Section~\ref{sec:discussion:falling}.

\begin{deluxetable}{rccc}
    \tablecaption{Global Polarization Spectrum. The median polarization fraction at each HAWC\texttt{+} band (53 \micron, 89 \micron, 154 \micron) normalized to the longest HAWC\texttt{+} band at 214 \micron.}
    \tablehead{\colhead{Model} & \colhead{$p_{53}/p_{214}$} & \colhead{$p_{89}/p_{214}$} & \colhead{$p_{154}/p_{214}$}}
    \startdata
		\homogeneous & $0.97 \pm 0.13$ & $0.99 \pm 0.05$ & $1.00 \pm 0.01$ \\ 
		\xiten & $0.94 \pm 0.15$ & $0.97 \pm 0.05$ & $0.99 \pm 0.01$ \\ 
		\zetahalf & $1.04 \pm 0.25$ & $1.01 \pm 0.08$ & $1.00 \pm 0.02$ \\ 
		\zetatwo & $1.29 \pm 0.33$ & $1.09 \pm 0.10$ & $1.02 \pm 0.02$ \\ 
		\hline
		OMC-1\tablenotemark{a} & $1.19 \pm 0.32$ & $1.19 \pm 0.21$ & $0.99 \pm 0.09$ \\ 
    \enddata
    \tablenotetext{a}{OMC-1 results from \citet{2021ApJ...907...46M}.}
    \label{tab:global_polspec}
\end{deluxetable}

\subsection{Pixel-by-Pixel Polarization Spectrum} \label{sec:analysisresults:pixelbypixel}
To investigate any effect local environments can have on the polarization spectrum, we look for quantitative variations in the polarization spectra computed individually for each pixel of our multi-wavelength synthetic polarization maps.
We fit these pixel-by-pixel polarization spectra to the form \citep{2016ApJ...824...84G, 2019ApJ...872..197S, 2021ApJ...907...46M}:
\begin{equation}\label{eq:linear_function}
    p \left( \lambda \right) / p \left( \lambda_0 \right) = a_\ell \left[ b_\ell \left( \lambda - \lambda_0 \right) + 1\right]
\end{equation}
As before, we use 214 $\mu$m as the normalizing wavelength (i.e.,~$\lambda_0 = 214$~$\mu\text{m}$). The shape of the polarization spectrum is thus characterized by the $b_\ell$ parameter.\footnote{The `true' slope is given by $a_\ell \times b_\ell$, but as $a_\ell\approx1$ in our results (see Table \ref{tab:median_fit_results_polspec}), we use $b_\ell$ as a proxy for the slope throughout this work.} A positive value for $b_\ell$ indicates a rising spectrum (i.e., polarization increasing with wavelength). A negative value for $b_\ell$ indicates a falling spectrum (i.e., polarization decreasing with wavelength). $a_\ell$ and $b_\ell$ are computed via a non-linear least squares fit for each pixel. For each set of synthetic observations, the median and MAD values for the resulting linear fit parameters are listed in Table~\ref{tab:median_fit_results_polspec}. We also include the results of the analogous analysis of OMC-1 in \citet{2021ApJ...907...46M}.
\begin{deluxetable}{rccc}
    \tablecaption{Median and median absolute deviation (MAD) values for the resulting pixel-by-pixel linear parameter fits (see Equation~\ref{eq:linear_function}).}
    \tablehead{\colhead{Model} & \colhead{$a_\ell$} & \colhead{$b_\ell \times 1000$ ($\mu\text{m}^{-1}$)} }
    \startdata
		\homogeneous & 	$1.00 \pm 0.02$ & $\phm{-}0.19  \pm 0.72$  \\ 
		\xiten	         & 	$1.01 \pm 0.02$ & $\phm{-}0.36  \pm 0.80$  \\ 
		\zetahalf	     & 	$0.99 \pm 0.03$ & $-0.18 \pm 1.39$ \\ 
		\zetatwo         & 	$0.96 \pm 0.05$ & $-1.65 \pm 1.95$ \\ 
		 \hline
		OMC-1\tablenotemark{a} &  $0.95\pm0.05$ & $-1.47\pm2.04$ \\
    \enddata
    \tablenotetext{a}{OMC-1 results from \citet{2021ApJ...907...46M}.}
    \label{tab:median_fit_results_polspec}
\end{deluxetable}
Overall, these results show consistency with the results of the global polarization spectrum analysis in Section~\ref{sec:analysisresults:global_spectrum}. Again, the homogeneous grain alignment model (\homogeneous) as well as the collisional depolarization grain alignment model~(\xiten) show flat or rising polarization spectra ($b_\ell \geq 0$). Negative values of $b_\ell$ are only seen for the temperature-dependent polarizability models with \zetatwo~again showing the most steeply falling spectrum that matches OMC-1. 

\begin{figure}
    \centering
    \includegraphics[width=0.45\textwidth]{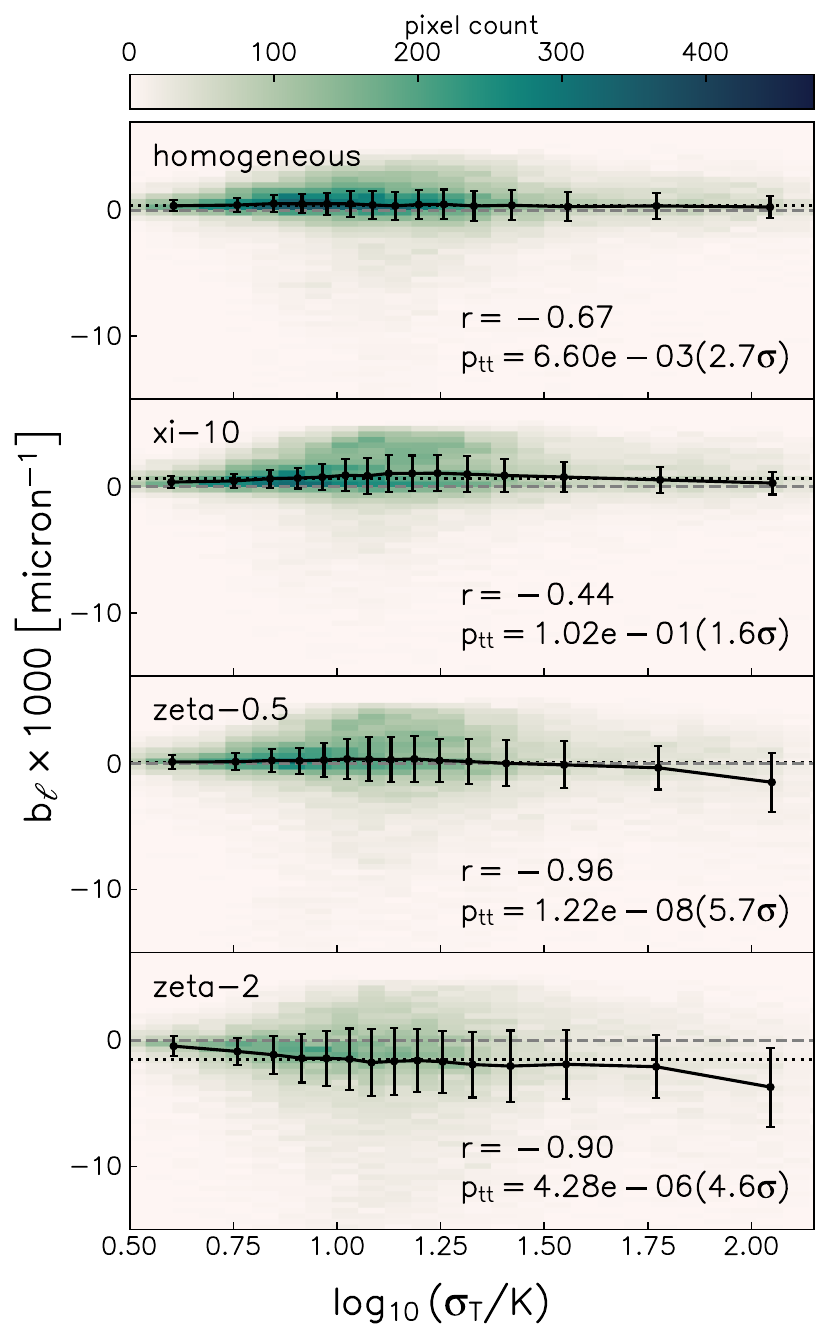}
    \caption{Correlation between slope of the polarization spectrum ($b_\ell$) and the standard deviation of the temperature along the line of sight ($\sigma_T$) for various grain alignment models. For each grain alignment model, each pixel is binned based on $b_\ell$ and $\sigma_T$. 
    The black line plots the median $b_\ell$ at 15 equally-sized bins in $\sigma_T$ with the errorbars indicating the median absolute deviation (MAD) at that $\sigma_T$ bin.
    The dashed grey line indicates $b_\ell = 0$ while the dotted black line shows the overall median $b_\ell$ of the grain alignment model.
    {The two-tailed Pearson correlation coefficient ($r$) and associated two-tailed p-values ($p_{tt}$) are also shown.}
    Only the temperature-dependent polarizability models show {statistically significant correlations} between $\sigma_T$ and $b_\ell$.
    }
    \label{fig:stdT_comparison}
\end{figure}

The HCE describes the scenario where a falling spectrum is achieved by the superposition of cooler, unaligned grains in dense regions along the line of sight with warmer, aligned grains in less dense regions. 
To assess whether the HCE is occurring in any of our models, we compare the slope of the polarization spectrum $b_\ell$ against the standard deviation of the voxel temperature along the line of sight ($\sigma_{\rm T}$) in Figure~\ref{fig:stdT_comparison}. 
These voxel temperature values are derived from the original simulation data. 
{We apply the two-tailed Pearson correlation coefficient test to compute the correlation coefficient ($r$) and associated two-tailed p-value ($p_{tt}$) on 15 equally-sized binned median values. These are also shown in Figure~\ref{fig:stdT_comparison}.}
The only grain alignment models to exhibit any {statistically significant} correlation between $b_\ell$ and  $\sigma_{\rm T}$ are the ones that include a temperature-dependent polarizability (i.e., \zetahalf, \zetatwo), which are also the only ones that a show falling spectrum. 
We discuss this further in Section~\ref{sec:discussion}.

\subsubsection{Correlations of the Polarization Spectrum}\label{sec:analysisresults:pixelbypixel:corr} 
\citet{2021ApJ...907...46M} explored the correlations between the polarization spectrum slope (i.e., $b_\ell$) and SED fitting derived temperature and column density maps obtained from \citet{2019ApJ...872..187C}. We apply this analysis to \zetatwo---the model that best matches the falling spectrum result of OMC-1.

\begin{figure*}
    \centering
    \includegraphics[width=0.98\textwidth]{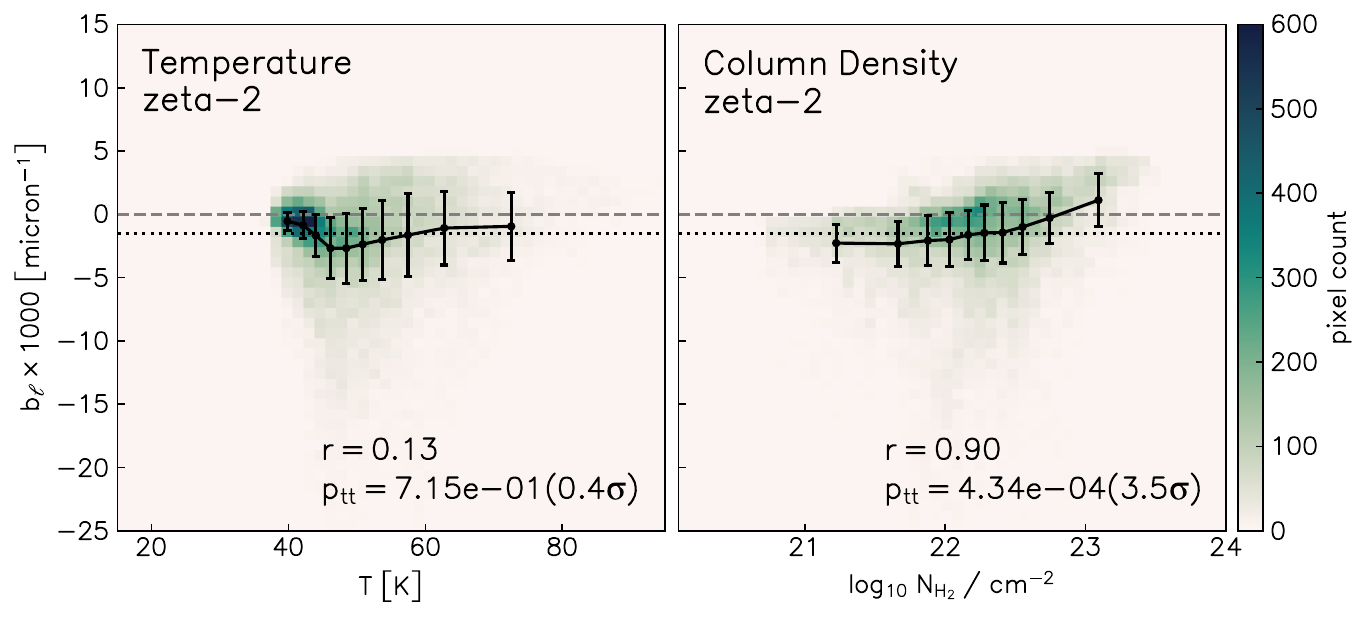}
    \caption{Correlations of $b_\ell$ with temperature (left) and column density (right) for the \zetatwo~model ($\alpha_p\propto T^2$). The black line indicates the median value of 10 equally-sized bins. The errorbars indicate the median absolute deviation value. The two-tailed Pearson correlation coefficient ($r$) and associated two-tailed p-values ($p_{tt}$) are also shown.
    The dashed grey line indicates $b_\ell = 0$ while the dotted black line indicates the overall median $b_\ell$ of the grain alignment model.
    \zetatwo, the grain alignment model tested that most closely matches the polarization spectrum of OMC-1, indicates no significant correlation between $b_\ell$ and temperature, but does show a significant correlation between $b_\ell$ and column density. See Section~\ref{sec:discussion:corr} for more discussions.
    }
    \label{fig:correlations_p_T_2}
\end{figure*}

Figure~\ref{fig:correlations_p_T_2} plots the slope of the polarization spectrum $b_\ell$ against the temperature and column density, respectively.
The figure also show the the median and MAD slope values for ten equally-sized bins. In addition, using the two-tailed Pearson correlation coefficient test, we calculate the correlation coefficient ($r$) and associated two-tailed p-value ($p_{tt}$) on the equally-sized binned median values. These statistics are shown in Figure~\ref{fig:correlations_p_T_2}. No robust correlation is found between $b_\ell$ and temperature. However, we do find evidence of a statistically significant correlation between $N_{\rm H_2}$ and $b_\ell$. We compute a Pearson correlation coefficient of $r=0.90$ with a significance of $3.5\sigma$.

{
We note that the dust model used to compute temperature and column density for OMC-1 differs from the one used in our analysis. 
We use the dust model from \citet{2001ApJ...548..296W}, whereas the model used for the observations of OMC-1 assumes a modified blackbody fit incorporating a dust emissivity index ($\beta$) as a free parameter \citep{2019ApJ...872..187C}.
Replicating our analysis with the model from \citet{2019ApJ...872..187C} does not alter any observed correlations (or lack thereof) between $N_{{\rm H}_2}$ and $b_\ell$ or between $T$ and $b_\ell$.
}



\section{Discussion}\label{sec:discussion}

\subsection{Origin of the Falling Polarization Spectrum}\label{sec:discussion:falling}
While observations of the far-infrared spectrum in dense molecular clouds have typically attributed its shape to the HCE \citep{1999ApJ...516..834H, 2008ApJ...679L..25V, 2013ApJ...773...29Z, 2021ApJ...907...46M}, there are a variety of different scenarios that could produce a falling far-infrared polarization spectrum.

In principle, it is possible for a falling spectrum to emerge without the consideration of any complexity in grain alignment or composition. For example, the falling spectrum (such as the one seen for OMC-1 in Figure~\ref{fig:all_polspec}) may arise purely as a consequence of magnetic field morphology. Cooler regions along the line of sight (traced well at longer wavelengths) tend to be denser and undergoing the process of gravitational collapse and star formation. The magnetic field in these regions may therefore be more tangled and disordered, thus resulting in a lower polarization fraction \citep[e.g.,][]{2016ApJ...829...84C}. On the other hand, the magnetic field of warmer regions along the line of sight (traced well at shorter wavelengths), without these ongoing processes, are likely more ordered. 
As a result, even if we assume the grain alignment efficiency is the same, the \textit{observed} polarization at longer wavelengths may be suppressed. 

In our model of homogeneous grain alignment, however, we find no evidence of a falling spectrum (see Figure~\ref{fig:all_polspec}a). As our simulations include variations in environmental conditions along the line-of-sight (as described in Section~\ref{sec:methods:simulations}), this implies that temperature variations, field geometry, etc.~along a sight line \textit{alone} do not generally result in a falling spectrum. In other words, the inability of the uniform grain alignment model to reproduce the falling polarization spectrum implies that variations in grain alignment efficiency are required to explain the falling spectrum.

The HCE posits that the falling spectrum originates from heterogeneous grain alignment efficiency: the line of sight includes both regions of cold unaligned grains and warmer regions of aligned grains. 
As cold regions can be cooler due to shielding from radiation, these colder regions are often denser.
One possible source of heterogeneous grain alignment{, for example, is} the effect of local density. High density regions facilitate more gas-grain collisions that serve to ``unalign'' grains \citep{2015ARA&A..53..501A}. 
{However, depolarization can also result from elevated temperatures disrupting grain alignment in regions of more modest densities.}

To evaluate the possibility of {such potentially competing} effects, we consider results from the grain alignment prescription from \citet{Lam_2021} as described in Section~\ref{sec:methods:grain_alignment_prescriptions}, 
By varying the value of the alignment parameter $\xi$ (i.e., \xione, \xihundred), we are able to explore the dependence of polarization on collisional dealignment. 
Nonetheless the falling spectrum expected of HCE is not observed (see Figure~\ref{fig:all_polspec}b and Figure~\ref{fig:all_polspec}c). 

As shown in Figure~\ref{fig:stdT_comparison}, we find evidence that the grain collisions are incapable of producing the HCE. Like the homogeneous grain model, it shows no correlation between $\sigma_T$---proxy for temperature heterogeneity along a line of sight---and the slope of the polarization spectrum.
For these collisional depolarization grain alignment models, there is no evidence that the superposition of different temperatures along the line of sight is causing a falling spectrum.
As such, we find that a collisional depolarization grain model is unable to produce a falling spectrum.

In Section~\ref{sec:methods:grain_alignment_prescriptions}, we constructed two models to emulate HCE by incorporating a relation between the emitted polarized emission at each voxel and local temperature: \zetahalf~and \zetatwo. As shown in Figure~\ref{fig:all_polspec}d, these two models represent the only models that are able to produce a falling spectrum. While both models produced a falling spectrum, it is the model that has a stronger dependence on temperature that delivers a polarization spectrum comparable to that observed in OMC-1:   $\alpha_p \propto T^2$ (\zetatwo).

Considering all the grain alignment models discussed in this work, the evidence implies that variations in grain alignment efficiency is required for a falling polarization spectrum (i.e., a homogeneous grain alignment model is insufficient). 
Additionally, we find that models that include collisional dealignment fail to produce a falling spectrum.
However, we do not rule out the possibility of other density-related effects.
For example, it is possible that grain growth or coagulation may also influence the shape of the polarization spectrum \citep{2013A&A...559A.133Y}. 
Nonetheless, we find that only our grain alignment model that is coupled to temperature is able to reproduce the falling far-infrared polarization spectrum.
This suggests that a falling spectrum and the HCE can be explained by RATs or similar mechanism where grain alignment is enhanced by ambient radiation sources.



Simulations from \citet{2007ApJ...663.1055B} do not produce a falling polarization spectrum despite including a realistic molecular cloud as well as a prescription for RATs. In comparing their polarization spectrum results with those from observations, \citet{2007ApJ...663.1055B} note that their simulations only include the interstellar radiation field (ISRF) and suggests that the existence of the falling spectrum is due to the presence of embedded sources---something not included in their simulation. The simulation used in our work does include embedded sources (the forming protostars; see Section~\ref{sec:methods:simulations}) suggesting that the presence of embedded radiation sources may indeed be essential for the falling polarization spectrum to be produced.

\subsection{Polarization Spectrum Correlations}
\label{sec:discussion:corr}
While we find a polarization spectrum comparable to that observed in OMC-1 when we use a simulation where the fractional polarization depends on temperature (i.e., \zetatwo), we do not find the same correlations that \citet{2021ApJ...907...46M} find between the slope of the polarization spectrum and temperature.
Specifically, for OMC-1, \citet{2021ApJ...907...46M} find a robust positive correlation between $b_\ell$ and $T$ ($r=0.93$, $p_{tt} = 6.5 \times 10^4$, $3.4\sigma$). We find no such correlation in our results between $b_\ell$ and $T$. On the other hand, while \citet{2021ApJ...907...46M} finds no correlation between the slope of the polarization spectrum and column density, our simulations show a statistically significant positive correlation between $b_\ell$ and $N_{{\rm H}_2}$ (i.e., regions of lower column density show more steeply falling spectrum). 

Taken together, \citet{2021ApJ...907...46M} present their correlations in OMC-1 as evidence for RATs as the explanation for HCE. The authors argue that it is the lack of radiation (as traced by lower dust temperature) that causes loss of alignment and not any intrinsic high column density effect. We can then ask why---despite exhibiting the HCE---do we not find similar correlations in our \zetatwo~($\alpha_{p}\propto T^2$) model?

\subsubsection{
\texorpdfstring{Correlations between $N_{{\rm H}_2}$ and $b_\ell$}{Correlations between Column Density and Slope}}\label{sec:discussion:corr_n}
A possible origin for the positive correlation between column density and $b_\ell$ is 
 wavelength-dependent polarization by absorption due to differences in optical depth at different wavelengths. 
The polarized emission from deep in the cloud can be absorbed by the aligned grains closer to the surface.
While this effect is generally negligible for clouds that are optically thin, it can be more pronounced at sufficiently high optical depths.
As optical depth increases at shorter wavelengths, this will have the result of lowering the observed polarization at the shorter wavelengths, effectively increasing the value of $b_\ell$ and thereby producing a more positive slope. 

To test this, we model the effect that optical depth induced absorption may have on the slope of the polarization. We follow the formalism from \citet{1989ApJ...345..802N} that calculates the magnitude of the optical depth induced reduction on the degree of polarization assuming a uniform magnetic field direction. 
Here, $p_{m, \lambda}$ is the measured polarization fraction at wavelength $\lambda$ while $p_{0,\lambda}$ represents the polarization fraction if there were no absorption. The emissivity at a given wavelength, $\varepsilon_\lambda$, is given by $\varepsilon_\lambda=1-\exp\left(-\tau_\lambda\right)$ for arbitrary optical depth ($\tau_\lambda$). To first order, we can approximate this effect as:
\begin{equation}\label{eq:novak1989firstorder}
    \frac{p_{m,\lambda}}{p_{0,\lambda}} = - \frac{1}{\varepsilon_\lambda} \left(1-\varepsilon_\lambda\right)\ln\left(1-\varepsilon_\lambda
    \right) 
\end{equation}

From our simulations for the four wavelength bands, we integrated along each sight line to obtain the optical depth using the dust model from \citet{2001ApJ...548..296W}. These optical depth values are then used to compute~$\varepsilon_\lambda$.
Assuming an initially flat polarization spectrum (i.e, $p_{0,53}=p_{0,89}=p_{0,154}=p_{0,214}$), we follow Equation~\ref{eq:novak1989firstorder} to estimate $p_{m,\lambda}$ and produce a model polarization spectrum.

\begin{figure}
    \centering
    \includegraphics[width=0.45\textwidth]{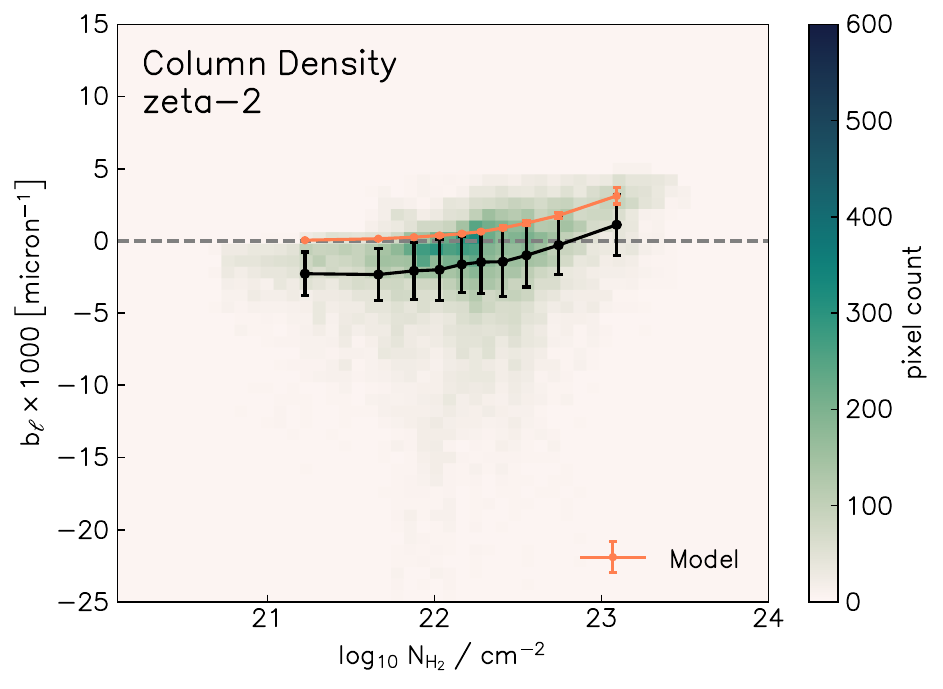}
    \caption{
    The correlation between $b_\ell$ and column density for the model \zetatwo~(shown in black) as in the right panel of Figure~\ref{fig:correlations_p_T_2}, overplotted with the optical depth model (shown in orange) adopted from \citet{1989ApJ...345..802N} (see Section~\ref{sec:discussion:corr_n}). 
    The dashed grey line indicates $b_\ell = 0$.
    The two $b_\ell-N_{\text{H}_2}$ curves are similar indicating that the $b_\ell-N_{\text{H}_2}$ correlation measured in our synthetic observation may be due to an optical depth effect.
    }
    \label{fig:tau_correction}
\end{figure}

To investigate how this model polarization spectrum correlates with column density, we compute $b_\ell$ and compare it to column density. This model is shown in Figure~\ref{fig:tau_correction} in orange. Like our simulation result, this model shows a positive correlation between $b_\ell$ and column density. While the orange curve is based on assuming an intrinsically flat spectrum, assuming an intrinsically falling or rising spectrum yields a similarly shaped curve. 
We emphasize the that model from \citet{1989ApJ...345..802N}, unlike our simulations, assumes a uniform magnetic field direction. As such, this comparison is only approximate. Nonetheless, we suggest that the correlations between $b_\ell$ and $N_{{\rm H}_2}$ can at least partially be attributed to this optical depth effect.


{Furthermore, we note that there is a slight rise in the value of $b_\ell$ for the highest column density bin of OMC-1 \citep[See Figure 4 in][]{2021ApJ...907...46M}. 
As such, we conclude that optical depth effects can influence the observed far-infrared polarization spectra of star forming clouds, and alter correlations of the spectrum properties with column density.}


\subsubsection{
\texorpdfstring{Correlations between $T$ and $b_\ell$}{Correlations between Temperature and Slope}
}
While our temperature-dependent polarizability grain alignment models show features of the HCE, they do not show a positive correlation between $b_\ell$ and $T$ (see Figure~\ref{fig:correlations_p_T_2}). 
{We do note that our simulated cloud is warmer when compared to OMC-1 (see Figure~\ref{fig:hist_comparison}) and that this is likely the result of the 35~K initial simulation temperature and temperature floor imposed by the simulation (detailed in Section~\ref{sec:methods:simulations}).
As such it is possible that the temperature floor is artificially affecting potential correlations. 
For example, a positive correlation may otherwise be visible should the simulation have had an initial gas temperature of less than 35~K.
While we cannot completely rule out all potential effects of this temperature difference on the correlations between $T$ and $b_\ell$, we note that only a small fraction of the simulation voxels are likely reset by this artificial temperature floor.}

{Regardless, we suggest} there is no reason to necessarily expect a relation between these two parameters for a cloud with the HCE. As described in \citet{1999ApJ...516..834H}, the crucial element of the HCE is the presence of both warmer and cooler components along the same sight line. The ``average'' or any single temperature determined for the sight line---such as that calculated from a SED fit---does not describe the presence of warmer and cooler components. By contrast, using $\sigma_{T}$ as described in Section~\ref{sec:analysisresults:pixelbypixel}, we can directly investigate the superposition of warmer and cooler components. As shown in Figure~\ref{fig:stdT_comparison}, we find that our \zetatwo~model shows a clear correlation between $b_\ell$ and the $\sigma_{T}$. This provides evidence that a correlation between $b_\ell$ and $T$ is not required for simulations that exhibit HCE.

\section{Conclusions}\label{sec:conclusison}

We generate synthetic far-infrared polarization spectra using a fully-3D, RMHD star-forming simulation and compare it with SOFIA/HAWC\texttt{+} observations of OMC-1 \citep{2021ApJ...907...46M}.  
By varying the grain alignment prescriptions used to generate the synthetic observations from our simulations, we investigated the origin of the far-infrared falling spectrum seen in observations of star-forming molecular clouds.

Our principal conclusions are as follows:
\begin{enumerate}

    \item We were not able to obtain a far-infrared falling polarization spectrum (negative slope of $p$ vs, $\lambda$) comparable to that observed in OMC-1 using a homogeneous grain alignment model. The polarization spectrum of our simulation using a homogeneous grain alignment model is flat.

    \item Similarly, with the inclusion of collisional depolarization through the grain alignment model of \citet{Lam_2021}, the polarization spectrum continues to be flat or rising. This remains the case along various viewing angles and for various values of the model parameter $\xi$ that is controlled by e.g., grain size and grain magnetic susceptibility.
    While collisional depolarization cannot produce a falling polarization spectrum, we note that other effects such as grain growth and coagulation \citep{2013A&A...559A.133Y} may also play a role and thus warrant further investigation.

    \item Extending this collisional depolarization grain alignment model by including a temperature-dependent polarizability does result in a falling polarization spectrum. 
    While these temperature-dependent polarizability models are not based on a realistic treatment of RATs and only represent a simple emulation of the expected feature of RATs, we find that both the \zetahalf~and \zetatwo~models are capable of producing a falling polarization spectrum. 
    The polarization spectrum of the \zetatwo~model, where $\alpha_p\propto T^2$, is most consistent with the OMC-1 observations. Furthermore, among the models evaluated here, the temperature-dependent polarizability models are the only ones that provide evidence for the heterogeneous cloud effect (HCE). Thus, HCE caused by RATs is a plausible explanation for the falling polarization spectrum. However, we note that the temperature dependent polarizability models used in this work represent a simple emulation of the features of RATs.
    As such, future work using more sophisticated models of grain alignment (such as \textsc{Polaris}; \citealt{2016A&A...593A..87R}) should be used to verify this relation.

    \item For the model that most closely matches OMC-1 (\zetatwo), we find a positive correlation between the column density and slope of the polarization spectrum. We find that this can be the result of optical depth induced depolarization, suggesting that optical depth effects can affect the observed far-infrared polarization spectra, and potentially alter correlations of the spectrum properties with column density.
    
    \item In contrast to \citet{2021ApJ...907...46M}, despite finding a similar overall falling spectrum in \zetatwo, we find no evidence of a correlation between temperature and the polarization spectrum slope. We suggest that this feature need not exist for HCE to explain the overall falling spectrum and may not be an expected general consequence from HCE. 

\end{enumerate}

In summary, our investigation of a series of grain alignment models in a realistic, fully-3D RMHD star forming molecular cloud---comparable to OMC-1---suggests that a falling polarization spectrum requires variations in grain alignment efficiency. While we find evidence that HCE caused by RATs can explain falling polarization spectrum we suggest further investigations with more sophisticated grain alignment models is required.

\section*{Acknowledgements}
This work is based on observations made with the NASA/DLR Stratospheric Observatory for Infrared Astronomy (SOFIA). SOFIA is jointly operated by the Universities Space Research Association, Inc. (USRA), under NASA contract NNA17BF53C, and the Deutsches SOFIA Institut (DSI) under DLR contract 50 OK 2002 to the University of Stuttgart. Financial support for this work was provided by NASA through award \#09-0535 to Villanova University issued by USRA.
The numerical simulation considered in this study was supported by NASA through a NASA ATP grant NNX17AK39G as well as the NASA High-End Computing (HEC) Program through the NASA Advanced Supercomputing (NAS) Division at Ames Research Center, and by the National Energy Research Scientific Computing Center (NERSC), a U.S. Department of Energy Office of Science User Facility located at Lawrence Berkeley National Laboratory, operated under Contract No.\,DE-AC02-05CH11231 using NERSC award NP-ERCAP0023065.
This work used computing resources provided by Northwestern University and the Center for Interdisciplinary Exploration and Research in Astrophysics (CIERA). This research was supported in part through the computational resources and staff contributions provided for the Quest high performance computing facility at Northwestern University which is jointly supported by the Office of the Provost, the Office for Research, and Northwestern University Information Technology.
This work was performed under the auspices of the U.S. Department of Energy (DOE) by Lawrence Livermore National Laboratory under Contract DE-AC52-07NA27344 (CYC). LLNL-JRNL-862957. 
{EGC acknowledges support from the the National Science Foundation MPS-Ascend Postdoctoral Research Fellowship under Grant No. 2213275.}
ZYL is supported in part by NSF AST-2307199 and NASA 80NSSC20K0533.

\vspace{5mm}
\facilities{SOFIA(HAWC\texttt{+})}


\software{\texttt{python}, \texttt{numpy} \citep{harris2020array}, \texttt{matplotlib}\citep{Hunter:2007}, \texttt{scipy} \citep{2020SciPy-NMeth}, \texttt{astropy} \citep{2013A&A...558A..33A,2018AJ....156..123A, 2022ApJ...935..167A}, \texttt{aplpy}}

\bibliography{bibliography, software}{}
\bibliographystyle{aasjournal}

\end{document}